\documentclass[twocolumn]{aa}
\usepackage{graphicx}
\usepackage{txfonts}
\newcommand{\flx}{\left< \dot{N}_{\mbox{\tiny Lyc}}^{\mbox{\tiny diffuse}} \right>}
\begin{document}

\title{The Formation of Free-Floating Brown Dwarves \\
and Planetary-Mass Objects \\
by Photo-Erosion of Prestellar Cores}

\author{A. P. Whitworth\inst{1} \and H. Zinnecker\inst{2}}

\offprints{A. P. Whitworth}

\institute{School of Physics \& Astronomy, Cardiff University, 
5 The Parade, Cardiff CF24 3YB, Wales, UK \\
\email{ant@astro.cf.ac.uk} \and
Astrophysikalisches Institut Potsdam, An der Sternwarte 16, D14482 Potsdam, 
Brandenburg, Germany \\
\email{hzinnecker@aip.de}}

\date{Received; accepted}

\abstract{
We explore the possibility that, in the vicinity of an OB star, a prestellar 
core which would otherwise have formed an intermediate or low-mass star may 
form a free-floating brown dwarf or planetary-mass object, because the outer 
layers of the core are eroded by the ionizing radiation from the OB star before 
they can accrete onto the protostar at the centre of the core. The masses of 
objects formed in this way are given approximately by $\sim 0.010 M_\odot\,
( a_{\rm I} / 0.3\,{\rm km}\,{\rm s}^{-1} )^6\,
( \dot{\cal N}_{\rm Lyc} / 10^{50}\,{\rm s}^{-1} )^{-1/3}\,
( n_{\rm 0} / 10^3\,{\rm cm}^{-3} )^{-1/3}\,$, where $a_{\rm I}$ 
is the isothermal sound speed in the neutral gas of the core, 
$\dot{\cal N}_{\rm Lyc}$ is the rate of emission of Lyman continuum photons 
from the OB star (or stars), and $n_{\rm 0}$ is the 
number-density of protons in the HII region surrounding the core. We 
conclude that the formation of low-mass objects by this 
mechanism should be quite routine, because the mechanism operates over 
a wide range of conditions ($10^{50}\,{\rm s}^{-1} \la \dot{\cal N}_{\rm Lyc} 
\la 10^{52}\,{\rm s}^{-1}$, $10\,{\rm cm}^{-3} \la n_0 \la 10^5\,{\rm cm}^{-3}$, 
$0.2\,{\rm km}\,{\rm s}^{-1} \la a_{\rm I} \la 0.6\,{\rm km}\,{\rm s}^{-1}$)
and is very effective. However, it is also a rather wasteful way of forming 
low-mass objects, in the sense that it requires a relatively massive initial 
core to form a single low-mass object. The effectiveness of photo-erosion 
also implies that that any intermediate-mass protostars which have 
formed in the vicinity of a group of OB stars must already have been well on 
the way to formation before the OB stars switched on their ionizing radiation; 
otherwise these protostars would have been stripped down to extremely low mass.

}

\titlerunning{Free-Floating Brown Dwarves and Planetary-Mass Objects}

\maketitle

\section{Introduction}

The existence of brown dwarves was first proposed on theoretical grounds by 
Kumar (1963a,b) and Hayashi \& Nakano (1963), 
but more than three decades passed before observations were able to establish 
their existence unambiguously (Rebolo, Zapatero Osorio \& Mart\'in, 1995; 
Nakajima et al., 1995; Oppenheimer et al., 1995). Now they are routinely 
observed (e.g. McCaughrean et al., 1995; Luhman et al., 1998; Wilking, 
Greene \& Meyer, 1999; Luhman \& Rieke, 1999; Lucas \& Roche 2000; Mart\'in 
et al. 2000; Luhman et al,, 2000; B\'ejar et al. 2001; Mart\'in et al., 2001; 
Wilking et al, 2002; McCaughrean et al., 2002). It appears that a 
significant fraction of brown dwarves in clusters are not companions 
to more massive stars, but rather they are free-floating. 
Reipurth \& Clarke (2001) have suggested that this is because brown dwarves 
are ejected from nascent multiple systems before they can acquire sufficient 
mass to become Main Sequence stars, and this suggestion is supported 
by the numerical simulations of Bate et al. (2002), Delgado Donate et al. (2003) 
and Goodwin et al. (2004a,b). Here we explore an 
alternative possibility, namely that in the central regions of clusters like the 
Orion Nebula Cluster, isolated brown dwarves can form from pre-existing prestellar 
cores which are eroded by the ambient ionizing radiation field. Similar 
explanations for the formation of free-floating low-mass objects 
have been proposed by Zapatero Osorio et al. (2000) and Kroupa (2002), but they 
do not appear to have been explored in detail before now.

If a dense neutral core is immersed in an HII region, the ambient Lyman 
continuum radiation drives an ionization front into the core. The ionization 
front is normally preceded by a compression wave, and this may render the 
core gravitationally unstable. At the same time, the flow of ionized gas off 
the core reduces its mass and injects material and kinetic energy into 
the surrounding HII region. This situation was analyzed by Dyson (1968; see 
also Kahn, 1969) who argued that it could explain the broadening and splitting 
of emission lines from the Orion Nebula. He also considered the circumstances 
under which such a core might become gravitationally unstable.

Bertoldi (1989) and Bertoldi \& McKee (1990) have developed semi-analytic 
models for larger cometary globules created 
by the interaction of ionizing radiation with a pre-existing cloud on the 
edge of an HII region. They use these semi-analytic models to evaluate the 
acceleration of the globule (due to the rocket effect; cf. Kahn, 1954; Oort \& 
Spitzer 1955) and its gravitational stability. Their results agree broadly 
with the numerical simulations of Sandford et al. (1982) and Lefloch \& 
Lazareff (1994). More recently Kessel-Deynet \& Burkert (2002) have 
modelled the radiation driven implosion of a pre-existing large cloud 
having internal structure. They use Smoothed Particle Hydrodynamics to 
handle the three-dimensional, self-gravitating gas dynamics, and find 
that small-scale internal structure delays, and may even inhibit, 
gravitational collapse.

However, in this paper we are concerned -- like Dyson and Kahn -- with 
small-scale neutral cores immersed well within an HII region, in the vicinity 
of the exciting stars, rather than large cometary globules at the edge of an 
HII region. In particular, we wish to evaluate the competition 
between erosion of material from the outer layers of such a core (due to 
the ionization front which eats into the core), and the formation and growth 
of a central protostar (triggered by the compression wave driven into the 
core ahead of the ionization front).

\section{Overview of Model}

In order to keep free parameters to a minimum in this exploratory 
analysis, we make a number of simplifying assumptions. We 
assume that there exists a prestellar core 
in equilibrium, and that a nearby massive star (or group thereof) then 
switches on its ionizing radiation instantaneously, and ionizes 
the surroundings of the prestellar core, on a time-scale much shorter 
than the sound-crossing time for the core (we justify this assumption 
later). The resulting increase in external pressure triggers the collapse 
of the core, {\it from the outside in}, and at the same time an ionization 
front starts to eat into the core, thereby eroding it from the outside, 
as it contracts. The final stellar mass is 
therefore determined by a competition between the rate of accretion 
onto the central protostar, and the rate of erosion at the boundary. 
Erosion ceases to be effective when the ionization front has eaten 
so far into the collapsing core that the newly ionized gas flowing 
off the ionization front is -- despite its outward kinetic energy -- 
still gravitational bound to the protostar.

The evolution consists of three phases, starting from the instant 
$t=0$ when the surroundings of the prestellar core first become ionized. 
At this juncture a compression wave is driven into the core setting 
up a uniform, subsonic inward velocity field. When this wave impinges 
on the centre at time $t_{\mbox{\tiny 1}}$, the first phase terminates and 
a central protostar is formed, which 
then grows steadily by accretion. At the same time an expansion 
wave is reflected and propagates outwards leaving an approximately 
freefall velocity field in its wake. At time $t_{\mbox{\tiny 2}}$ this outward 
propagating expansion wave encounters the inward propagating 
ionization front, and the second phase terminates. Thereafter the ionization 
front is eroding material 
which is falling ever faster inwards. Eventually, at time $t_{\mbox{\tiny 3}}$, 
the ionization front encounters material which is falling inwards so fast 
that it cannot be unbound by the ionization front, and the third phase 
terminates. The final stellar mass 
is the mass interior to the ionization front at time $t_{\mbox{\tiny 3}}$.

In Section 3 we derive an expression for the diffuse ionizing 
radiation field in an HII region, and in Section 4 we formulate 
the speed with which the ionization front eats into the core. 
In Section 5, we describe the initial prestellar core and formulate 
the time $t_{\mbox{\tiny 1}}$ it takes for the compression wave to propagate to 
its centre. In Section 6 we describe the modified density and 
velocity field set up in the inner core by the outward propagating 
expansion wave, and formulate the time $t_{\mbox{\tiny 2}}$ at which 
the outward propagating expansion wave  meets the inward propagating 
ionization front. In Section 7 we formulate the time $t_{\mbox{\tiny 3}}$ 
at which the ionization front has eaten so far into the modified density 
field inside the expansion wave that it has encountered material 
which is infalling too fast to be unbound by the act of ionization. 
The material interior to the ionization front at time $t_{\mbox{\tiny 3}}$ is 
presumed to constitute the final star. We can therefore obtain an 
expression for its mass.

\section{The diffuse ionizing flux}

We shall assume that the HII Region which engulfs the pre-existing 
prestellar core is excited by a compact cluster of OB stars which 
emits Lyman continuum (i.e. hydrogen ionizing) photons at a constant 
rate $\dot{\cal N}_{\mbox{\tiny Lyc}}$. We shall also assume that most of the 
surrounding space is occupied by a uniform gas in which the density 
of hydrogen nuclei in all forms is $n_{\mbox{\tiny 0}}$. The radius of the 
initial HII Region is therefore
\begin{equation} \label{EQN:RHII}
R_{\mbox{\tiny HII}} \;=\; \left( \frac{3\,\dot{\cal N}_{\mbox{\tiny Lyc}}}
{4\,\pi\,\alpha_*\,n_{\mbox{\tiny 0}}^2} \right)^{1/3}\,,
\end{equation}
where $\alpha_* \simeq 2 \times 10^{-13}\,{\rm cm}^3\,{\rm s}^{-1}$ 
is the recombination coefficient into excited states only; in other 
words, we are using the On-The-Spot Approximation.

If we neglect the shadows cast by other prestellar cores and attenuation by 
dust inside the HII region, the radial -- or direct -- number-flux of Lyman 
continuum photons $\dot{N}_{\mbox{\tiny Lyc}}^{\mbox{\tiny radial}}$ (i.e. 
the number of ionizing photons crossing unit area in unit time) varies with 
distance $R$ from the OB cluster as
\begin{equation}
\dot{N}_{\mbox{\tiny Lyc}}^{\mbox{\tiny radial}} \;=\; 
\frac{\dot{\cal N}_{\mbox{\tiny Lyc}}}
{4\,\pi\,R^2} \,-\, \frac{\alpha_*\,n_{\mbox{\tiny 0}}^2\,R}{3}\,.
\end{equation}
where the first term on the righthand side represents geometric dilution and 
the second term represents the absorptions which maintain ionization against 
recombination. Hence the mean value of $\dot{N}_{\mbox{\tiny Lyc}}^
{\mbox{\tiny radial}}$, averaged over the volume of the HII region, is
\begin{equation}
\label{RADIAL}
\left< \dot{N}_{\mbox{\tiny Lyc}}^{\mbox{\tiny radial}} \right> \;\simeq\; 
\frac{3}{4}\,\left( \frac{3\,\dot{\cal N}_{\mbox{\tiny Lyc}}\,\alpha_*^2\,
n_{\mbox{\tiny 0}}^4}{4\,\pi} \right)^{1/3}\,.
\end{equation}

The volume number-emissivity for the diffuse component of the Lyman continuum 
radiation field (i.e. the number of ionizing photons emitted from unit volume 
in unit time into unit solid angle, due to recombinations straight into the 
ground state) is
\begin{equation}
\dot{n}_{\mbox{\tiny Lyc}}^{\mbox{\tiny diffuse}} \;=\; \frac{\alpha_{_1}
\,n_{\mbox{\tiny 0}}^2}{4\,\pi}\,,
\end{equation}
where $\alpha_{_1} \simeq 10^{-13}\,{\rm cm}^3\,{\rm s}^{-1}$ is 
the recombination coefficient into the ground state only. The 
mean diffuse flux is therefore of order
\begin{equation}
\label{DIFFUSE}
\left< \dot{N}_{\mbox{\tiny Lyc}}^{\mbox{\tiny diffuse}} \right> \;\sim\;
\pi\,\dot{n}_{\mbox{\tiny Lyc}}\,R_{\mbox{\tiny HII}} \;\simeq\; 
\frac{\alpha_{_1}}{4\,\alpha_*}\,\left( \frac{3\,
\dot{\cal N}_{\mbox{\tiny Lyc}}\,\alpha_{_*}^2\,n_{\mbox{\tiny 0}}^4}
{4\,\pi} \right)^{1/3}\,.
\end{equation}
Comparing Eqns. (\ref{RADIAL}) and (\ref{DIFFUSE}), we see that 
the mean diffuse flux is smaller than the mean radial flux by a 
factor $\sim \alpha_{_1} / 3 \alpha_* \sim 1/6$. This is to some 
extent compensated by the fact that a spherical core of 
radius $r$ presents a larger 
effective cross-section to the diffuse flux ($4 \pi r^2$) than 
to the radial flux ($\pi r^2$), i.e. the net dose of Lyman 
continuum radiation received by a spherical core divides roughly 
in the ratio 60\% 
radial and 40\% diffuse. In order to proceed with a spherically 
symmetric analysis of this phenomenon, we shall assume that the 
flux of Lyman continuum radiation incident on a core is 
constant, isotropic, and equal to $2\left< \dot{N}_{\mbox{\tiny 
Lyc}}^{\mbox{\tiny diffuse}} \right>\,$. 
However, we should be mindful that this is a crude representation 
of the Lyman continuum radiation field incident on a real core in 
a real HII region. We will check retrospectively 
that the cores we invoke are much smaller than the HII region in 
which they are imersed (see Section 8).

\section{The photo-erosion rate}

We denote the instantaneous radius of the neutral core by 
$r_{\mbox{\tiny IF}}$, and the density of hydrogen nuclei in all 
forms in the neutral core by 
$n_{\mbox{\tiny I}}(r),\;r\leq r_{\mbox{\tiny IF}}\,$, 
where $r$ is distance measured from the centre of the core. We 
assume that the ionization front eats into the core at speed 
$\left[ -\,\dot{r}_{\mbox{\tiny IF}} \right]\,<\,a_{\mbox{\tiny I}}\,$, 
where $a_{\mbox{\tiny I}}$ is the isothermal sound-speed in the neutral gas. 
Consequently the ionization front is not preceded by a strong shock front, 
just a weak compression wave. This is a reasonable assumption, except near 
the beginning, when the ionization front advances very rapidly into the core. 
We can only allow for this by performing numerical simulations. However, this 
initial phase of rapid erosion does not last long, and will not greatly 
affect the overall phenomenology of core erosion or the final protostellar 
mass. It will simply shorten the time interval, $t_{_{1}}$, before the 
protostar forms. Thus the expressions for the final protostellar mass which 
we derive later (Eqns. \ref{EQN:APPROXMASS} and \ref{EQN:EFFICIENCY}) should 
be taken as indicative rather than precise.

Conservation of mass across the ionization front 
requires 
\begin{equation}
\label{EQN:MASSCON1}
n_{\mbox{\tiny I}}(r_{\mbox{\tiny IF}})\,
\left[ -\dot{r}_{\mbox{\tiny IF}} \right] \;=\; 
n_{\mbox{\tiny II}}(r_{\mbox{\tiny IF}})\,
\left[ v_{\mbox{\tiny II}}(r_{\mbox{\tiny IF}})-
\dot{r}_{\mbox{\tiny IF}} \right] \,,
\end{equation}
where $n_{\mbox{\tiny II}}(r_{\mbox{\tiny IF}})$ and $v_{\mbox{\tiny II}
}(r_{\mbox{\tiny IF}})$ are -- respectively -- the density of hydrogen 
nuclei in all forms and the radial speed of the ionized gas, just outside 
the neutral core.

To proceed, we assume that the ionized gas flows away from the ionization 
front at constant speed
\begin{equation} \label{EQN:OUTFLOW}
\left[ v_{\mbox{\tiny II}}(r) \,-\, \dot{r}_{\mbox{\tiny IF}} \right] \;=\; 
\left[ v_{\mbox{\tiny II}}(r_{\mbox{\tiny IF}}) \,-\, \dot{r}_{\mbox{\tiny IF}} 
\right] \;=\; a_{\mbox{\tiny II}} \,, 
\end{equation}
where $a_{\mbox{\tiny II}}$ is the isothermal sound-speed in the ionized gas. 
(In reality (Kahn 1969) the flow of ionized gas downstream from the ionization 
front accelerates away from the core due to the pressure gradient, but 
this only changes the result we derive below by a numerical factor of order 
unity.) Hence the density in the ionized gas falls off approximately as 
$r^{-2}$, 
\begin{equation}
n_{\mbox{\tiny II}}(r) \;=\; n_{\mbox{\tiny II}}(r_{\mbox{\tiny IF}})\,
\left( \frac{r}{r_{\mbox{\tiny IF}}} \right)^{-2}\,,
\end{equation}
until it approaches the background density $n_{\mbox{\tiny 0}}$ at radius 
$r_{_\infty} \gg r_{\mbox{\tiny IF}}$. Also, substituting from Eqn. 
\ref{EQN:OUTFLOW} into Eqn. \ref{EQN:MASSCON1} gives
\begin{equation} \label{EQN:MASSCON2}
n_{\mbox{\tiny I}}(r_{\mbox{\tiny IF}})\,\left[ -\dot{r}_{\mbox{\tiny IF}} \right]
 \;\simeq\; n_{\mbox{\tiny II}}(r_{\mbox{\tiny IF}})\,a_{\mbox{\tiny II}}\,.
\end{equation}

Almost all the incident flux of Lyman continuum photons is used up 
balancing recombinations in the outflowing ionized gas, so we can 
put
\begin{eqnarray}\label{EQN:JACKET}
2\flx & \simeq & \int_{r=r_{\mbox{\tiny IF}}}^{r=r_{_\infty}} \alpha_*\,
n_{\mbox{\tiny II}}^2(r)\,dr \;\simeq\;\frac{\alpha_*\,n_{\mbox{\tiny II}}^2
(r_{\mbox{\tiny IF}})\,r_{\mbox{\tiny IF}}}{3} \,; \\ \label{EQN:nIIrIF}
n_{\mbox{\tiny II}}(r_{\mbox{\tiny IF}}) & \simeq & \left( \frac{6\,\flx}{\alpha_*\,
r_{\mbox{\tiny IF}}} \right)^{1/2} \,; \\ \label{RDOT}
\dot{r}_{\mbox{\tiny IF}} & \simeq & -\,\frac{n_{\mbox{\tiny II}}
(r_{\mbox{\tiny IF}})\,a_{\mbox{\tiny II}}}{n_{\mbox{\tiny I}}
(r_{\mbox{\tiny IF}})} \; \simeq \; - \, \left( 
\frac{6\,\flx}{\alpha_*\,r_{\mbox{\tiny IF}}} \right)^{1/2} \, 
\frac{a_{\mbox{\tiny II}}}{n_{\mbox{\tiny I}}(r_{\mbox{\tiny IF}})} \,.
\end{eqnarray}
In obtaining the final expression in Eqn. (\ref{EQN:JACKET}), we have 
assumed that $r_{_\infty} \gg r_{\mbox{\tiny IF}}$. Again this is a 
reasonable assumption except near the beginning.

\section{The first phase ($0 < t < t_{\mbox{\tiny 1}}$)}

\subsection{The initial core}

For the purpose of this exploratory calculation, we assume that 
the pre-existing prestellar core is a singular isothermal sphere. 
Generalizing the treatment to a non-singular (i.e. Bonnor-Ebert) 
isothermal sphere should not change the result greatly, but would 
necessitate a detailed numerical formulation of the problem; it 
would tend to reduce the mass of the final protostar, and hence 
assist in producing low-mass final objects.

In a singular isothermal sphere, the density of hydrogen nuclei 
in all forms, $n_{\mbox{\tiny I}}(r)$  and the mass interior to radius $r$, 
$M(r)$, are given by
\begin{eqnarray}\label{STATIC}
n_{\mbox{\tiny I}}(r) & = & \frac{a_{\mbox{\tiny I}}^2}{2\,\pi\,G\,m\,r^2}\,, \\
M(r) & = & \frac{2\,a_{\mbox{\tiny I}}^2\,r}{G}\,,
\end{eqnarray}
where $m = m_{\mbox{\tiny p}}/X = 2.4 \times 10^{-24}\,{\rm g}$ is the 
mass associated with one hydrogen nucleus when account is taken 
of other elements (in particular helium), $m_{\mbox{\tiny p}}$ is the 
mass of a proton, and $X$ is the fractional abundance of hydrogen 
by mass. If the boundary density is $n_{\mbox{\tiny 0}}$, the outer radius 
and total mass are initially
\begin{eqnarray} \label{EQN:CORERADIUS}
r_{\mbox{\tiny 0}} & = & \frac{a_{\mbox{\tiny I}}}{\left( 2\,\pi\,G\,
n_{\mbox{\tiny 0}}\,m \right)^{1/2}}\,, \\ \label{EQN:COREMASS}
M_{\mbox{\tiny 0}} & = & \left( \frac{2}{\pi\,G^3\,n_{\mbox{\tiny 0}}\,m} 
\right)^{1/2}\,a_{\mbox{\tiny I}}^3\,.
\end{eqnarray}

\subsection{The inward propagating compression wave}

When this core is first overrun by the HII region, at time $t = 0$, 
the resulting increase in the external pressure drives a 
compression wave into the core, thereby triggering its collapse. 
We assume that the compression wave travels ahead of the ionization front. 
It leaves in its wake a subsonic -- and approximately uniform -- 
inward velocity field. The compression wave reaches the centre 
at time
\begin{equation}
t_{\mbox{\tiny 1}} \;=\; \frac{r_{\mbox{\tiny 0}}}
{a_{\mbox{\tiny I}}} \;=\; \frac{1}{\left( 2\,
\pi\,G\,n_{\mbox{\tiny 0}}\,m \right)^{1/2}}\,.
\end{equation}
i.e. after one sound-crossing time. We can now justify the assumption that 
the time-scale on which the HII region overruns the core is much shorter 
than a sound-crossing time. The time-scale for the establishment of the 
Initial Str$\o$mgren Sphere is of order the recombination time-scale, viz.
\[
t_{\mbox{\tiny recombination}} \sim \left( \alpha_* 
n_{\mbox{\tiny 0}} \right)^{-1} \,,
\]
so the condition becomes $t_{\mbox{\tiny recombination}} \ll 
t_{\mbox{\tiny 1}}\,$, or
\[
n_{\mbox{\tiny 0}} \;\gg\; \frac{2\,\pi\,G\,m}{\alpha_*^2} \;\sim\; 
3 \times 10^{-5} \,{\rm cm}^{-3}\,.
\]
This is easily satisfied since typically $n_{\mbox{\tiny 0}} 
\stackrel{>}{\sim} 1\,{\rm cm}^{-3}\,$. 

\section{The second phase ($t_{\mbox{\tiny 1}} < t < t_{\mbox{\tiny 2}}$)}

\subsection{Formation of the central protostar}

At $t = t_{\mbox{\tiny 1}}$ a central protostar forms and subsequently grows in 
mass at a constant rate
\begin{equation} \label{EQN:ACCNRATE}
\dot{M}_* \;\simeq\; \frac{a_{\mbox{\tiny I}}^3}{G}\,.
\end{equation}
We note that this result is true for the collapse of a singular isothermal 
sphere irrespective of whether its collapse is triggered -- as here -- {\it 
from outside-in} by the inward propagation of a mild compression wave (see 
Whitworth \& Summers, 1985), or -- as in the Shu (1977) model -- {\it from 
inside-out} by a central perturbation.

\subsection{The expansion wave}

At the same time that the compression wave converges on the centre, 
an expansion wave is launched outwards at speed 
$a_{\mbox{\tiny I}}$, relative to the gas. If we neglect the small 
inward velocity already acquired by the gas due to the inward propagating 
compression wave, the radius of the expansion wave is given by
\begin{equation}
\label{EXPWAVE}
r_{_{\mbox{\tiny EW}}} \;\simeq\; a_{\mbox{\tiny I}}\,
\left( t-t_{\mbox{\tiny 1}} \right) \;\simeq\; 
a_{\mbox{\tiny I}}t \,-\, r_{\mbox{\tiny 0}} \,.
\end{equation}
We now introduce dimensionless variables
\begin{eqnarray}
\xi & = & \frac{r}{r_{\mbox{\tiny 0}}}\,, \\
\tau & = & \frac{t}{t_{\mbox{\tiny 1}}}\,,
\end{eqnarray}
in terms of which the radius of the expansion wave is given by
\begin{equation}
\label{EXPWAVE_DIM}
\xi_{_{\mbox{\tiny EW}}} \;=\; \tau \,-\, 1 \,, \hspace{1cm} \tau \;>\; 1 \,.
\end{equation}

\subsection{The core interior to the expansion wave}

Interior to the expansion wave, material flows inwards, to feed 
the constant accretion rate onto the central protostar, and quickly 
approaches a freefall-like profile. Therefore the density and velocity 
can be approximated by
\begin{eqnarray}
\left. \begin{array}{rcl}\label{EQN:INFALL}
n'_{\mbox{\tiny I}}(r) & = & \left( 3 / 8 \pi G m \right)\,
a_{_{\mbox{\tiny I}}}^2\,r_{_{\mbox{\tiny EW}}}^{-1/2}\,r^{-3/2} \,, \\
 & & \\
v'_{_{\mbox{\tiny I}}}(r) & = & -\,\left( 2/3 \right)\,a_{_{\mbox{\tiny I}}}\,
r_{_{\mbox{\tiny EW}}}^{1/2}\,r^{-1/2} \,, 
\end{array} \; \right\} \;\; t>t_{_{\mbox{\tiny 1}}}\,,\,\;
r<r_{_{\mbox{\tiny EW}}}\,.
\end{eqnarray}
There is a degree of freedom in specifying the coefficients of 
$n'_{\mbox{\tiny I}}(r)$ and $v'_{_{\mbox{\tiny I}}}(r)$. We have chosen 
them so that (i) they are consistent with the constant accretion rate 
(see Eqn. \ref{EQN:ACCNRATE}), and (ii) the mass of the central protostar 
and the mass in the accretion flow interior to the expansion wave are 
equal. Formally the velocity changes discontinuously across the expansion 
wave, from zero outside to $-2a_{\mbox{\tiny I}}/3$ inside; in reality 
the change will be much smaller since the inward propagating compression 
wave during the first phase will have set up an inward velocity of comparable 
magnitude (e.g. Whitworth \& Summers 1985).

\subsection{The ionization front exterior to the expansion wave}

Substituting for $n_{\mbox{\tiny I}}(r_{\mbox{\tiny IF}})$ in Eqn. (\ref{RDOT}) 
from Eqn. (\ref{STATIC}), we obtain
\begin{equation}\label{IF1}
\frac{dr_{_{\mbox{\tiny IF}}}}{dt} \; = \; - \, \left( \frac{6\,\flx}{\alpha_* 
r_{_{\mbox{\tiny IF}}}} \right)^{1/2} \, \frac{2\,\pi\,G\,m\,a_{\mbox{\tiny II}}\,
r_{_{\mbox{\tiny IF}}}^2}{a_{\mbox{\tiny I}}^2}\,.
\end{equation}

If we now introduce a third dimensionless variable 
\begin{eqnarray} \nonumber
\beta & = & \left( \frac{6\,\flx\,r_{\mbox{\tiny 0}}^3}{\alpha_*} \right)
^{1/2} \, \frac{\pi\,G\,m\,a_{\mbox{\tiny II}}}{a_{\mbox{\tiny I}}^3} \\ \label{EQN:BETA}
 & = & \left( \frac{\pi\,3^8\,\alpha_{\mbox{\tiny 1}}^6\,G^3\,m^3\,
\dot{\cal N}_{\mbox{\tiny Lyc}}^2}{2^7\,\alpha_*^8\,a_{\mbox{\tiny I}}^6\,
n_{\mbox{\tiny 0}}} \right)^{1/12} \, \frac{a_{\mbox{\tiny II}}}
{2\,a_{\mbox{\tiny I}}} \,.
\end{eqnarray}
Eqn. (\ref{IF1}) reduces to
\begin{equation}
\frac{d\xi_{_{\mbox{\tiny IF}}}}{d\tau} \; = \; -\,2\,\beta \,
\xi_{_{\mbox{\tiny IF}}}^{3/2}\,,
\end{equation}
with solution
\begin{equation} \label{IF1_DIM}
\xi_{_{\mbox{\tiny IF}}} \;=\; \left[ 1\,+\,\beta \tau \right]^{-2}\,.
\end{equation}
To fix the constant of integration, we have used the initial 
condition, {\it viz.}: at $t=0$, $r_{_{\mbox{\tiny IF}}} = 
\xi_{\mbox{\tiny IF}}r_{\mbox{\tiny 0}} = r_{\mbox{\tiny 0}}$; 
and hence at $\tau = 0$, $\xi_{_{\mbox{\tiny IF}}} = 1$.

\subsection{The ionization front meets the expansion wave}

The ionization front meets the expansion wave at time $t_{\mbox{\tiny 2}}$ 
when $r_{\mbox{\tiny IF}}=r_{\mbox{\tiny EW}}$, or in terms of the 
dimensionless variables, at $\tau_{\mbox{\tiny 2}}$ when 
$\xi_{\mbox{\tiny IF}} = \xi_{\mbox{\tiny EW}} = \xi_{\mbox{\tiny 2}}$. 
Equating Eqns. (\ref{IF1_DIM}) and (\ref{EXPWAVE_DIM}), we obtain
\begin{equation}\label{BLOB}
\left( 1\,+\,\beta\tau_{\mbox{\tiny 2}} \right)^{-2} \;=\; 
\tau_{\mbox{\tiny 2}}\,-\,1 \,.
\end{equation}

With typical values $\dot{\cal N}_{\mbox{\tiny Lyc}} = 10^{50}\,
{\rm s}^{-1}$ and $n_{\mbox{\tiny 0}} = 10^4\,{\rm cm}^{-3}$, we have 
$\flx \simeq 3 \times 10^{11}\,{\rm cm}^{-2}\,{\rm s}^{-1}$. 
Putting $a_{\mbox{\tiny I}} = 0.3\,{\rm km}\,{\rm s}^{-1}$ and $a_{\mbox{\tiny II}} 
= 10\,{\rm km}\,{\rm s}^{-1}$, we obtain $\beta \simeq 25$. Hence, 
if we put
\begin{equation}
\xi_{\mbox{\tiny 2}} \;=\; \tau_{\mbox{\tiny 2}}\,-\,1\,,
\end{equation}
it follows that $\xi_{\mbox{\tiny 2}} \ll 1$, and Eqn.(\ref{BLOB}) becomes 
\begin{equation}\label{EQN:MEETING}
\left[ (1+\beta) \,+\, \beta \xi_{\mbox{\tiny 2}} \right]^2\,
\xi_{\mbox{\tiny 2}} \;=\; 1 \,,
\end{equation}
with approximate solution
\begin{equation} \label{EQN:MASS2}
\xi_{\mbox{\tiny 2}} \;\simeq\; \left( 1\,+\,\beta \right)^{-2} \;\sim\; 0.002 \,.
\end{equation}
This is the dimensionless radius of the inward propagating ionization front 
when it meets the outward propagating expansion wave.

\begin{figure*}
\label{LOCI}
\setlength{\unitlength}{1mm}
\begin{picture}(50,140)
\includegraphics{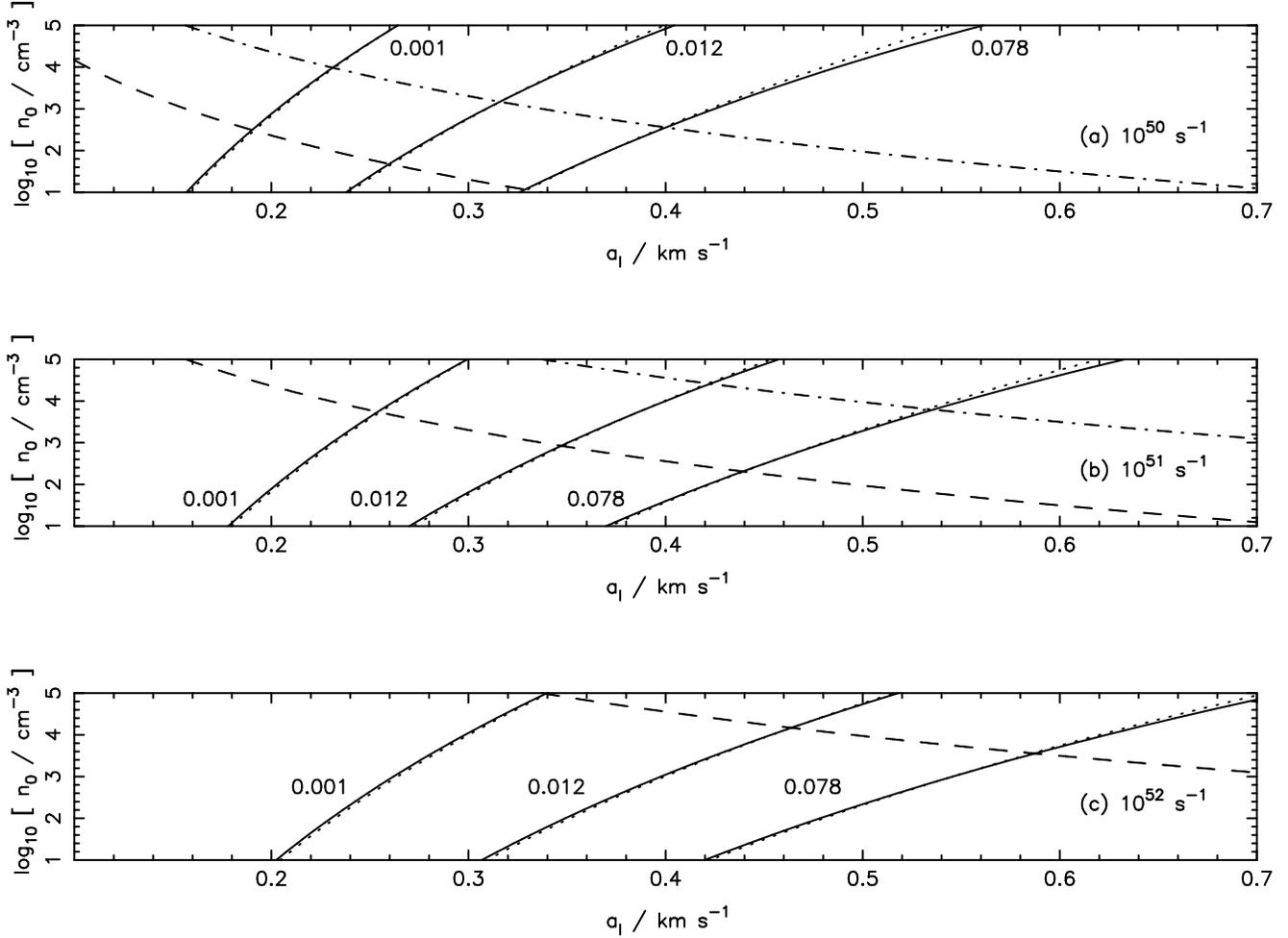}
\end{picture}
\caption{For (a) $\dot{\cal N}_{\rm Lyc} = 10^{50}\,{\rm s}^{-1}$, (b) $\dot{\cal N}_
{\rm Lyc} = 10^{51}\,{\rm s}^{-1}$, and (c) $\dot{\cal N}_{\rm Lyc} = 10^{52}\,{\rm s}
^{-1}$, the solid lines show the loci on the $(a_{\rm I},\ell og_{\rm 10}[n_0])$-plane
where free-floating objects of mass $0.001\,M_\odot$ (one Jupiter mass), $0.012\,
M_\odot$ (the deuterium-burning limit) and $0.078\,M_\odot$ (the hydrogen-burning 
limit) form due to photo-erosion of pre-existing cores. The dotted lines show the 
predictions of the approximate Eqn. \ref{EQN:APPROXMASS}. The dashed (dash-dot) line 
is the locus below which the volume of the initial core is less than 0.001 (0.01) of 
the volume of the whole HII region.}
\end{figure*}

\section{The third phase ($t_{\mbox{\tiny 2}} < t < t_{\mbox{\tiny 3}}$)}

\subsection{The ionization front interior to the expansion wave}

Once the inward propagating ionization front passes the outward 
propagating expansion wave, the ionization front encounters 
infalling gas with density and velocity given by Eqn. \ref{EQN:INFALL}. 
Eqn. \ref{EQN:MASSCON2} therefore has to be adjusted to account for the 
inward motion of the neutral gas:
\begin{equation} \label{EQN:MASSCON3}
n'_{\mbox{\tiny I}}(r_{\mbox{\tiny IF}})\,\left[ -\dot{r}_{\mbox{\tiny IF}} 
\,+\, v'_{\mbox{\tiny I}}(r_{\mbox{\tiny IF}}) \right] \;\simeq\; 
n_{\mbox{\tiny II}}(r_{\mbox{\tiny IF}})\,a_{\mbox{\tiny II}}\,.
\end{equation}
Substituting for $n'_{\mbox{\tiny I}}(r_{\mbox{\tiny IF}})$ and 
$v'_{\mbox{\tiny I}}(r_{\mbox{\tiny IF}})$ from Eqn. \ref{EQN:INFALL} and for 
$n_{\mbox{\tiny II}}(r_{\mbox{\tiny IF}})$ from Eqn. \ref{EQN:nIIrIF}, we obtain
\begin{equation} \label{EQN:XIDOTIF}
\dot{\xi}_{\mbox{\tiny IF}} \;=\; -\,\frac{2\,(\tau - 1)^{1/2}}{3}\,\left[ 
\xi_{\mbox{\tiny IF}}^{-1/2} \,+\, 4\,\beta\,\xi_{\mbox{\tiny IF}} \right] \,.
\end{equation}

The solution is
\begin{equation} \label{EQN:TAUMINONE}
(\tau - 1)^{3/2} \;=\; \xi_{\mbox{\tiny 2}}^{3/2} \,+\, \frac{3}{8\,\beta}\,\ell n 
\left[ \frac{1\,+\, 4\,\beta\,\xi_{\mbox{\tiny 2}}^{3/2}}
{1\,+\, 4\,\beta\,\xi_{\mbox{\tiny IF}}^{3/2}} \right] \,,
\end{equation}
where we have fixed the constant of integration by requiring that 
$r_{\mbox{\tiny IF}} = r_{\mbox{\tiny EW}} = r_{\mbox{\tiny 2}}$ at 
$t_{\mbox{\tiny 2}}$; or equivalently $\xi_{\mbox{\tiny IF}} = 
\xi_{\mbox{\tiny EW}} = \xi_{\mbox{\tiny 2}} = \tau_{\mbox{\tiny 2}} - 1$ 
at $\tau_{\mbox{\tiny 2}}$.

To follow the ionization front into the third phase, we pick ever smaller values 
of $\xi_{\mbox{\tiny IF}}\;(<\xi_{\mbox{\tiny 2}})$ and calculate $\tau$ from Eqn. 
\ref{EQN:TAUMINONE}. The radial velocity of the ionization front is then given by
$\dot{r}_{\mbox{\tiny IF}} = a_{\mbox{\tiny I}} \dot{\xi}_{\mbox{\tiny IF}}$, where 
$\dot{\xi}_{\mbox{\tiny IF}}$ can be calculated from Eqn. \ref{EQN:XIDOTIF}. The 
specific kinetic energy of the newly ionized gas, relative to the central protostar 
is therefore
\begin{equation} \label{EQN:SPECKIN}
\epsilon_{\mbox{\tiny KIN}} \;=\; \frac{1}{2}\,\left[ \dot{r}_{\mbox{\tiny IF}} 
\,+\, a_{\mbox{\tiny II}} \right]^2 \;=\; \frac{a^2_{\mbox{\tiny I}}}{2}\,\left[ 
\dot{\xi}_{\mbox{\tiny IF}} \,+\, \frac{a_{\mbox{\tiny II}}}{a_{\mbox{\tiny I}}} 
\right]^2 \,.
\end{equation}

Taking into account the mass of the protostar, $M_* = G^{-1} a^2_{\mbox{\tiny I}} 
r_{\mbox{\tiny 0}} (\tau - 1)$, plus the mass in the accretion flow interior to 
the ionization front, $M_{\mbox{\tiny I}} = G^{-1} a^2_{\mbox{\tiny I}}
r_{\mbox{\tiny 0}} (\tau - 1)^{-1/2} \xi_{\mbox{\tiny IF}}^{3/2}$, the specific 
binding energy of this material is
\begin{equation} \label{EQN:SPECBIND}
\epsilon_{\mbox{\tiny BIND}} \;=\; \frac{G(M_* + M_{\mbox{\tiny I}})}
{r_{\mbox{\tiny IF}}} \;=\; a^2_{\mbox{\tiny I}}\,\left\{ \frac{(\tau - 1)}
{\xi_{\mbox{\tiny IF}}} \,+\, \frac{\xi_{\mbox{\tiny IF}}^{1/2}}
{(\tau - 1)^{1/2}} \right\} \,.
\end{equation}
We shall assume that ionization ceases to erode the core once the specific 
kinetic energy, $\epsilon_{\mbox{\tiny KIN}}$ (Eqn. \ref{EQN:SPECKIN}), is less 
than the specific binding energy, $\epsilon_{\mbox{\tiny BIND}}$ (Eqn. 
\ref{EQN:SPECBIND}). The mass $M_{\mbox{\tiny 3}} = M_* + M_{\mbox{\tiny I}}$ 
interior to the ionization front at this stage is the final mass of the protostar. 
This occurs at time $t_{\mbox{\tiny 3}} = \tau_{\mbox{\tiny 3}} t_{\mbox{\tiny 1}}$, 
where 
\begin{equation} \label{EQN:TAU3}
\frac{1}{2}\,\left[ \dot{\xi}_{\mbox{\tiny IF}}(\tau_{\mbox{\tiny 3}}) \,+\, 
\frac{a_{\mbox{\tiny II}}}{a_{\mbox{\tiny I}}} \right]^2  \;=\; 
\frac{(\tau_{\mbox{\tiny 3}} - 1)}{\xi_{\mbox{\tiny 3}}} \,+\, 
\frac{\xi_{\mbox{\tiny 3}}^{1/2}}{(\tau_{\mbox{\tiny 3}} - 1)^{1/2}} \,;
\end{equation}
and the final mass of the protostar is then given by
\begin{equation} \label{EQN:FINALMASS}
\frac{M_{\mbox{\tiny 3}}}{M_{\mbox{\tiny 0}}} \;=\; \frac{1}{2}\,\left[ 
(\tau_{\mbox{\tiny 3}} - 1) \,+\, (\tau_{\mbox{\tiny 3}} - 1)^{-1/2} 
\xi_{\mbox{\tiny 3}}^{3/2} \right] \,.
\end{equation}

\section{Results}

The procedure for obtaining the final protostellar mass $M_{\mbox{\tiny 3}}$ 
is therefore as follows. (i) Chose $(\dot{\cal N}_{\mbox{\tiny Lyc}}, 
a_{\mbox{\tiny I}}, 
n_{\mbox{\tiny 0}})$ and calculate $\beta$ from Eqn \ref{EQN:BETA}. (ii) Solve 
Eqn. \ref{EQN:MEETING} for $\xi_{\mbox{\tiny 2}}$. (iii) Solve Eqn. 
\ref{EQN:TAUMINONE} for $\xi_{\mbox{\tiny IF}}(\tau)$, ($\tau > 1 + 
\xi_{\mbox{\tiny 2}}$). (iv) Identify $\tau_{\mbox{\tiny 3}}$ satisfying 
Eqn. \ref{EQN:TAU3}. (v) Compute $M_{\mbox{\tiny 3}}$ according to Eqn. 
\ref{EQN:FINALMASS}.

Figure 1 shows, for three representative values of $\dot{\cal N}_{\mbox{\tiny Lyc}}$ 
[i.e. (a) 
$10^{50}\,{\rm s}^{-1}$, (b) $10^{51}\,{\rm s}^{-1}$, and (c) $10^{52}\,{\rm s}^{-1}$] 
loci of contant $M_{\mbox{\tiny 3}} = 0.001M_\odot,\;0.012 M_\odot,\;{\rm and}\;0.078 
M_\odot$ on the $(a_{\rm I},\ell og_{\rm 10}[n_0])$-plane. These values of 
$M_{\mbox{\tiny 3}}$ are chosen because they represent, respectively, one Jupiter 
mass, the deuterium-burning limit, and the hydrogen-burning limit. Notionally objects 
in the mass range $0.012 M_\odot \la M_{\mbox{\tiny 3}} \la 0.078 M_\odot$ are 
referred to as brown dwarves, whilst those in the mass range $M_{\mbox{\tiny 3}} 
\la 0.012 M_\odot$ are referred to as planetary-mass objects. We see that for a 
broad range of values of $\dot{\cal N}_{\mbox{\tiny Lyc}}$, $n_{\mbox{\tiny 0}}$ and 
$a_{\mbox{\tiny I}}$, there exists the possibility to form free-floating brown 
dwarves and planetary-mass objects, provided suitable pre-existing cores are 
available to be photo-eroded.

In estimating the radiation field incident on a core, we have assumed that 
the volume occupied by the core is much smaller than the volume of the HII region 
around it (see Section 3). It is therefore necessary to check that this is the case. 
If we require the ratio between the initial volume of a core and the volume of the 
whole HII region to be less than some fraction $\eta$, then substituting from 
Eqns. \ref{EQN:RHII} and \ref{EQN:CORERADIUS} we obtain the condition
\begin{equation}
n_0 \;<\; n_{\rm max}(\eta) \;=\; \frac{\pi}{2} 
\left( \frac{3\,\dot{\cal N}_{\mbox{\tiny Lyc}}\,\eta}{\alpha_*} \right)^2 
\left( \frac{G\,m}{a_{\mbox{\tiny I}}^2} \right)^3 \,.
\end{equation}
$n_{\rm max}(\eta)$ is plotted on Figure 1 for $\eta = 0.001$ (dashed curve) and
$\eta = 0.01$ (dash-dot curve). Evidently this is a significant constraint for 
$\dot{\cal N}_{\mbox{\tiny Lyc}} \sim 10^{50}\,{\rm s}^{-1}$, but for higher 
values of $\dot{\cal N}_{\mbox{\tiny Lyc}}$ it is less important.

Since the protostar does not grow hugely during the Third Phase, 
we can approximate the results by adopting the mass at the end of the 
Second Phase. If in addition we assume $\beta \gg 1$, so that from Eqn. 
\ref{EQN:MASS2} $\;\xi_{\mbox{\tiny 2}} \simeq \beta^{-2}$, we can write
\begin{eqnarray} \nonumber
M_{\mbox{\tiny 3}} & \simeq & f\,\dot{M}_*\,t_{\mbox{\tiny 1}}\,\xi_{\mbox{\tiny 2}} 
\\ \nonumber
 & \simeq & \frac{f\,2^3\,\alpha_{_*}\,a_{\mbox{\tiny I}}^6}{3\,\alpha_{_1}\,
a_{\mbox{\tiny II}}^2\,G^2\,m}\,\left( \frac{\alpha_{_*}}{6\,\pi^2\,
\dot{\cal N}_{\mbox{\tiny Lyc}}\,n_{\mbox{\tiny 0}}} \right)^{1/3} \\ \label{EQN:APPROXMASS}
 & \simeq & 0.01 M_\odot
\left( \frac{a_{\mbox{\tiny I}}}{0.3\,{\rm km}\,{\rm s}^{-1}} \right)^6
\left( \frac{\dot{\cal N}_{\mbox{\tiny Lyc}}}{10^{50}\,{\rm s}^{-1}} \right)^{-1/3}
\left( \frac{n_{\mbox{\tiny 0}}}{10^3{\rm cm}^{-3}} \right)^{-1/3}.
\end{eqnarray}
Here $f \simeq 1.7$ is a factor to account for the additional matter which accretes 
onto the protostar after the ionization fron meets the expansion wave. We can 
understand this factor as follows. In a collapsing singular isothermal sphere, the 
infalling mass interior to the expansion wave is equal to the mass already in the 
central protostar. Thus, in the problem considered here, when the ionization front 
meets the expansion wave at $t=t_{\mbox{\tiny 3}}$, the infalling mass interior to 
the ionization front equals the mass already in the protostar. It follows that if 
no further mass acretes onto the protostar, $f = 1$; and if all the infalling mass 
interior to the ionization front at $t=t_{\mbox{\tiny 3}}$ accretes onto the 
protostar, then $f = 2$. The empirical value $f = 1.7$ suggests that most of the 
infalling mass interior to the ionization front at $t=t_{\mbox{\tiny 3}}$ is 
accreted onto the protostar.

Eqn. \ref{EQN:APPROXMASS} is plotted as a dotted line on Figure 1. We see that 
it is a very good approximation to the numerical results. Given the approximate 
nature of the whole analysis, Eqn. \ref{EQN:APPROXMASS} can be viewed as our best 
estimate of the final protostellar mass.

\section{Discussion}

This exploratory analysis suggests that it is possible for a relatively 
massive prestellar core which suddenly finds itself immersed in an HII 
region to be stripped down to the mass of a brown dwarf, or even a planet, 
by photo-erosion, before it can collapse to form a star. To improve on this 
analysis, it is probably necessary to resort to numerical modelling. One 
could then, in principle,  
(i) use a more realistic configuration for the pre-existing prestellar 
core; (ii) include the development of the HII region as it overruns the 
core; (iii) treat properly the early phase when the ionization front 
advances very rapidly into the outer layers of the core, and the 
compression wave which is driven into the core; (iv) attempt to capture 
effects due to departures from spherical symmetry, in particular the radiative 
transfer aspects; (v) include self-gravity explicitly in the equations of 
motion. We suspect that the photo-erosion mechanism may be somewhat more effective 
than our analysis suggests, because firstly, with a prestellar core which is not 
as centrally condensed as a singular isothermal sphere (e.g. Ward-Thompson, 
Motte \& Andr\'e, 1999; Bacmann et al. 2000), the protostar will 
not form so quickly, giving the ionization front more time to erode; 
and secondly, the ionization front will penetrate at higher speed into the 
less dense central regions.

We can express the final protostellar mass $M_{\mbox{\tiny 3}}$ in terms of 
the initial core mass $M_{\mbox{\tiny 0}}$, by combining Eqns. 
\ref{EQN:APPROXMASS}, \ref{EQN:COREMASS} and \ref{EQN:RHII} to obtain
\begin{equation} \label{EQN:EFFICIENCY}
M_{\mbox{\tiny 3}} \simeq \frac{2\,f\,\alpha_{_*}\,G\,M_{\mbox{\tiny 0}}^2}
{3\,\alpha_{\mbox{\tiny 1}}\,a_{\mbox{\tiny II}}^2\,R_{\mbox{\tiny HII}}} \,.
\end{equation}
There are two important deductions which we can make from this equation.

First, the ratio of the final protostellar mass to the initial core mass is 
given by
\begin{equation}
\frac{M_{\mbox{\tiny 3}}}{M_{\mbox{\tiny 0}}} \;\sim\; 10^{-4} 
\left( \frac{M_{\mbox{\tiny 0}}}{M_\odot} \right) 
\left( \frac{R_{\mbox{\tiny HII}}}{{\rm pc}} \right)^{-1}\,.
\end{equation}
The photo-erosion process is therefore very effective, in the sense that 
it strips off most of the mass of the initial core. This makes it a rather 
wasteful way of creating low-mass 
protostars, in the sense that it requires a relatively massive initial core to 
create a single final low-mass protostellar object. In addition, we infer 
that any intermediate-mass protostars which have formed in the vicinity of a group 
of OB stars must already have been well on the way to formation before the OB 
stars switched on their ionizing radiation; otherwise these protostars would 
have been stripped down to extremely low mass.

Second, if the number of cores in the mass interval 
$(M_{\mbox{\tiny 0}},M_{\mbox{\tiny 0}}+dM_{\mbox{\tiny 0}})$ 
(i.e. the core mass function) has the form
\begin{equation}
{\cal N}_{M_{\mbox{\tiny 0}}}\,dM_{\mbox{\tiny 0}} \;\propto\; 
M_{\mbox{\tiny 0}}^{-\gamma}\,dM_{\mbox{\tiny 0}} \,,
\end{equation}
then the initial mass function for brown dwarves and planetary-mass objects 
formed by photo-erosion has the form
\begin{equation}
{\cal N}_{M_{\mbox{\tiny 3}}}\,dM_{\mbox{\tiny 3}} \;\propto\; 
M_{\mbox{\tiny 3}}^{-(\gamma +1)/2}\,dM_{\mbox{\tiny 3}} \,.
\end{equation}
Motte et al. (1998) find $\gamma \simeq 1.5$ for $M_{\mbox{\tiny 0}} < 0.5 M_\odot$ 
and $\gamma \simeq 2.5$ for $M_{\mbox{\tiny 0}} > 0.5 M_\odot$ for the cores in 
$\rho$ Ophiuchus, and similar results are reported for the cores in Ophiuchus by 
Johnstone et al. (2000), for the cores in Serpens by Testi \& Sargent (1998), and 
for the cores in the Orion B molecular cloud by Motte et al. (2001) and Johnstone 
et al. (2001). Hence the exponent in the initial mass function for brown dwarves 
and planetary-mass objetcs formed by photo-erosion should lie in the range 
$\,-\,1.25\,$ to $\,-\,1.75\,$, with the smaller value ($-\,1.75$) obtaining for 
higher-mass cores. As the HII region expands 
and engulfs additional cores, $R_{\mbox{\tiny HII}}$ increases, and therefore for 
fixed $M_{\mbox{\tiny 3}}$ the progenitor cores have larger masses $M_{\mbox{\tiny 0}}$. 
Consequently, as one moves away from the stars exciting the HII region, the density 
of brown dwarves and planetary-mass objects formed by photo-erosion should decrease, 
and their initial mass function should become slightly steeper.

We have omitted from our analysis the possible effect of a stellar wind 
impinging on a core (A Burkert, private communication). A simple estimate 
suggests that a stellar wind would need to be very powerful to have 
a significant effect. Suppose that the central OB star blows a wind with 
mass-loss rate $\dot{M}_{\rm wind} \sim 10^{-4}\,M_\odot\,{\rm year}^{-1}$, 
and speed $v_{\rm wind} \sim 1000\,{\rm km}\,{\rm s}^{-1}$. The resultant 
ram pressure acting on a core at radius $R$ is 
\begin{equation} \label{Pramwind}
P_{\rm wind} \;=\; \frac{\dot{M}_{\rm wind}\,v_{\rm wind}}{4\,\pi\,R^2} \,.
\end{equation}
We can compare this with the ram pressure of the ionized gas flowing off 
the surface of a core which is exposed to the direct flux of ionizing radiation:
\begin{equation} \label{PramIF}
P_{\rm IF} \;=\; \frac{\dot{\cal N}_{\rm Lyc}\,m\,a_{\rm II}}{4\,\pi\,R^2} \,.
\end{equation}
Substituting typical values, we obtain
\begin{eqnarray} \nonumber
\frac{P_{\rm wind}}{P_{\rm IF}} & = & \frac{\dot{M}_{\rm wind}\,v_{\rm wind}}
{\dot{\cal N}_{\rm Lyc}\,m\,a_{\rm II}} \;=\; 6 \times 10^{-3} \;
\left( \frac{\dot{M}_{\rm wind}}{10^{-4}\,M_\odot\,{\rm year}^{-1}} 
\right) \times \\
 & & \hspace{2cm} 
\left( \frac{v_{\rm wind}}{1000\,{\rm km}\,{\rm s}^{-1}} \right) \;
\left( \frac{\dot{\cal N}_{\rm Lyc}}{5 \times 10^{49}\,{\rm s}^{-1}} \right)^{-1} \,.
\end{eqnarray}
This implies that the wind can have little effect. In reality, most of the 
ionizing flux from the central star is spent maintaining ionization against 
recombination in the flow of ionized gas from the core, rather than producing 
new ionizations. Therefore Eqn. (\ref{PramIF}) significantly overestimates 
the ram pressure of the ionized gas flowing off the core, but probably not 
by two orders of magnitude. The effect of a stellar wind will be 
to modify the rate at which the ionized gas flowing off the core can disperse 
(e.g. Richling \& Yorke, 1998; Hollenbach, Yorke \& Johnstone, 2000). 
A bow shock will form on the side of the core facing into the wind (the 
windward side), and the ionized gas flowing off this side of the core will 
be swept back in the direction of the wind. This will have the effect of 
increasing the density, and hence the recombination rate, in the ionized 
gas flowing off the core, thereby reducing the rate of erosion somewhat. 
On the leeward side of the core the effect of the wind will be minimal.

\section{Conclusions}

The analysis we have presented in this paper is based on several  
assumptions and approximations, which, whilst reasonable, are  
idealizations of what is likely to occur in nature. With this proviso, 
we infer that, in the central regions of large clusters, free-floating 
brown dwarves and planetary-mass objects may form when more massive 
pre-existing cores are overrun by the HII regions excited by newly-formed 
massive stars. The sudden ionization of the surroundings of 
a core drives a compression wave into the core, creating at its 
centre a protostar which then grows by accretion. At the same time an 
ionization front starts eating into the core, thereby removing its 
outer layers. The final mass of the protostar is determined by a 
competition between the rate at which it can accrete the infalling envelope, and 
the rate at which the ionization front can erode the envelope. The simple 
analysis presented here suggests that in large, dense clusters this will 
result in the formation of free-floating brown dwarves and/or planetary-mass 
objects. The process is both robust, in the sense that it operates 
over a wide range of conditions, and also very effective, in the sense that 
it strips off most of the mass of the initial core. This means that it is a 
wasteful way of creating low-mass protostars, because a relatively massive 
initial core is required to produce a single final brown dwarf or 
planetary-mass object. It also means that any intermediate-mass protostars which 
have formed in the vicinity of a group of OB stars must already have been 
well on the way to formation before the OB stars switched on their ionizing 
radiation; otherwise these protostars would have been stripped down to 
extremely low mass.

\begin{acknowledgements}

We thank the referee, Doug Johnstone, who drew our attention to a fundamental 
error in the original version of this paper, and made several other important 
suggestions which we have incorporated in the final version. We gratefully 
acknowledge the support of a European Commission Research Training Network, 
awarded under the Fifth Framework (Ref. HPRN-CT-2000-00155).

\end{acknowledgements}


\begin{thebibliography}{}

\bibitem[2000]{bacmann} Bacmann, A., Andr\'e, P., Puget, J.-L., Abergel, A.,  Bontemps, S., Ward-Thompson, D., 2000, \aap, 361, 555

\bibitem[2001]{bate} Bate, M.R., Bonnell, I.A., Bromm, V., 2002, \mnras, 332 L65

\bibitem[2001]{bejar} B\'ejar, V.J.S., Mart\'in, E.L., Zapatero Osorio, M.R., 
Rebolo, R., Barrado y Navascu\'es, D., Bailer-Jones, C.A.L., Mundt, R., 
Baraffe, I., Chabrier, C., Allard, F., 2001, \apj, 556, 830

\bibitem[1989]{bertoldi1} Bertoldi, F., 1989, \apj, 346, 735

\bibitem[1990]{bertoldi2} Bertoldi, F., McKee, C.F., 1990, \apj, 354, 529

\bibitem[2003]{delgado} Delgado-Donate, E., Clarke, C. J., Bate, M. R., 2003, 
\mnras, 342, 926

\bibitem[1968]{dyson} Dyson, J.E., 1968, \apss, 1, 388

\bibitem[2004a]{goodwin1} Goodwin, S. P., Whitworth, A. P., Ward-Thompson, D., 
2004a, \aap, 414, 633

\bibitem[2004b]{goodwin2} Goodwin, S. P., Whitworth, A. P., Ward-Thompson, D., 
2004b, \aap, in press

\bibitem[1963]{hayashi} Hayashi, C., Nakano, T., 1963, Prog. Theor. Phys., 30, 460

\bibitem[2000]{hollenbach} Hollenbach, D.J., Yorke, H.W., Johnstone, D., 
2000, in {\it Protostars and Planets, IV} (University of Arizona Press, 
Tucson) Eds. V Mannings, AP Boss, SS Russell, 401

\bibitem[2000]{johnstone1} Johnstone, D., Wilson, C. D., Moriarty-Schieven, G., 
Joncas, G., Smith, G., Gregersen, E., Fich, M., 2000, \apj, 545, 327

\bibitem[2000]{johnstone2} Johnstone, D., Fich, M., Mitchell, G. F., 
Moriarty-Schieven, G., 2001, \apj, 559, 307

\bibitem[1954]{kahn1} Kahn, F.D., 1954, Bull. astron. Inst. Netherlands, 12, 187

\bibitem[1969]{kahn2} Kahn, F.D., 1969, Physica, 41, 172

\bibitem[2002]{kessel} Kessel-Deynet, O., Burkert, A., 2002, \mnras, 338, 545

\bibitem[2002]{Kroupa} Kroupa, P., 2002, Science, 295, 82

\bibitem[1963a]{Kumar1} Kumar, S.S., 1963a, \apj, 137, 1121

\bibitem[1963b]{Kumar2} Kumar, S.S., 1963b, \apj, 137, 1126

\bibitem[1994]{Lefloch2} Lefloch, B., Lazareff, B., 1994, \aap, 289, 559

\bibitem[2000]{Lucas} Lucas, P.W., Roche, P.F., 2000, \mnras, 314, 858

\bibitem[1998]{Luhman1} Luhman, K.L., Rieke, G.H., Lada, C.J., Lada, E.A., 
1998, \apj, 508, 347

\bibitem[1999]{Luhman2} Luhman, K.L., Rieke, G.H., 1999, \apj, 525, 440

\bibitem[2000]{Luhman3} Luhman, K.L., Rieke, G.H., Young, E.T., Cotera, A.S., 
Chen, H., Rieke, M.J., Schneider, G., Thompson, R.I., 2000, \apj, 540, 1016

\bibitem[1995]{McCaughrean95} McCaughrean, M.J., Zinnecker, H., Rayner, J.T., 
Stauffer, J., 1995, in {\it The bottom of the Main Sequence - and beyond} 
(ESO Astrophysics Symposium, Springer Verlag) Ed. C. G. Tinney, 209

\bibitem[2002]{McCaughrean02} McCaughrean, M. J., Zinnecker, H., Andersen, M., 
Meeus, G., Lodieu, N., 2002, The Messenger, 109, 28

\bibitem[2000]{Martin1} Mart\'in, E.L., Brandner, W., Bouvier, J., 
Luhman, K.L., Stauffer, J., Basri, G., Zapatero Osorio, M.R., Barrado y 
Navascu\'es, D., 2000, \apj, 543, 299

\bibitem[2001]{Martin2} Mart\'in, E.L., Zapatero Osorio, M.R., Barrado y 
Navascu\'es, D., B\'ejar, V.J.S., Rebolo, R., 2001, \apj, 558, L117

\bibitem[1998]{Motte1} Motte, F., Andr\'e, P., Neri, R., 1998, \aap, 336, 150

\bibitem[2001]{Motte2} Motte, F., Andr\'e, P., Ward-Thompson, D., Bontemps, S., 
\aap, 372, L41

\bibitem[1995]{Nakajima} Nakajima, T., Oppenheimer, B.R., Kulkarni, S.R., 
Golimowski, D.A., Matthews, K., Durrance, S.T., 1995, Nature, 378, 463

\bibitem[1955]{Oort} Oort, J.H., Spitzer, L., 1955, \apj, 121, 6

\bibitem[1995]{Oppenheimer} Oppenheimer, B.R., Kulkarni, S.R., Matthews, K., 
Nakajima, T., 1995, Science, 270, 1478

\bibitem[1995]{Rebolo} Rebolo, R., Zapatero Osorio, M.R., Mart\'in, E.L., 
1995, Nature, 377, 129

\bibitem[2001]{Reipurth} Reipurth, B., Clarke, C.J., 2001, \aj, 122, 432

\bibitem[1998]{Richling} Richling, S., Yorke, H.W., 1998, \aap, 340, 508

\bibitem[1982]{Sandford} Sandford, M.T. II, Whitaker, R.W., Klein, R.I., 
1982, \apj, 260, 183

\bibitem[1977]{Shu} Shu, F.H., 1977, \apj, 214, 488

\bibitem[1998]{Testi} Testi, L., Sargent, A.I., 1998, \apj, 508, L91

\bibitem[1999]{WardThompson} Ward-Thompson, D., Motte, F., Andr\'e, P., 
1999, \mnras, 305, 143

\bibitem[1985]{Whitworth} Whitworth, A.P., Summers, D., 1985, \mnras, 214, 1

\bibitem[1999]{Wilking1} Wilking, B.A., Greene, T.P., Meyer, M.R., 1999, 
\aj, 117, 469

\bibitem[2002]{Wilking2} Wilking, B.A., Mikhail, A., Carlson, G., Meyer, M.R., 
Greene, T.P., 2002, \aj, 127, 1131

\bibitem[2000]{Osorio} Zapatero Osorio, M.R., B\'ejar, V.J.S., Mart\'in, E.L., 
Rebolo, R., Barrado y Navascu\'es, D., Bailer-Jones, C.A.L., Mundt, R., 2000, 
Science, 290, 103

\end{thebibliography}
\end{document}